\documentclass[number,dvips]{arxstspdf}
\usepackage{graphicx}
\usepackage{flushend}
\usepackage{stfloats}
\usepackage{breakurl}

\volume{25}
\issue{4}
\pubyear{2010}
\firstpage{458}
\lastpage{475}
\doi{10.1214/09-STS281}

\makeatletter
\newproclaim{algorithm}{Algorithm}
\makeatother

\begin{document}
\begin{frontmatter}

\title{Learning Active Basis Models by EM-Type Algorithms}
\runtitle{Learning Active Basis Models by EM-Type Algorithms}

\begin{aug}
\author[a]{\fnms{Zhangzhang} \snm{Si}%
\ead[label=e1,text=zzsi@stat.ucla.edu]{zzsi@stat.ucla.edu}},
\author[a,b]{\fnms{Haifeng} \snm{Gong}%
\ead[label=e2,text=hfgong@stat.ucla.edu]{hfgong@stat.ucla.edu}},
\author[a,b]{\fnms{Song-Chun} \snm{Zhu}%
\ead[label=e3,text=sczhu@stat.ucla.edu]{sczhu@stat.ucla.edu}}\and
\author[a]{\fnms{Ying Nian} \snm{Wu}\corref{}%
\ead[label=e4,text=ywu@stat.ucla.edu]{ywu@stat.ucla.edu}}
\runauthor{Si, Gong, Zhu and Wu}

\affiliation{University of California and Lotus Hill Research Institute}

\address[a]{Zhangzhang Si is Ph.D. Student,
Department of Statistics, University of California, Los Angeles, USA
\printead{e1}.}
\address[b]{Haifeng Gong is Postdoctoral Researcher,
Department of Statistics, University of California, Los Angeles, USA
and Lotus Hill Research Institute, Ezhou, China
\printead{e2}.}
\address[c]{Song-Chun Zhu is Professor,
Department of Statistics, University of California, Los Angeles, USA
and Lotus Hill Research Institute, Ezhou, China
\printead{e3}.}
\address[d]{Ying Nian Wu is Professor,
Department of Statistics, University of California, Los Angeles, USA
\printead{e4}.}

\end{aug}

%
\begin{abstract}
EM algorithm is a convenient tool for maximum likelihood model
fitting when the data are incomplete or when there are latent
variables or hidden states. In this review article we explain
that EM algorithm is a natural computational scheme for learning
image templates of object categories where the learning is not fully
supervised. We represent an image template by an active basis model,
which is a linear composition of a selected set of localized, elongated
and oriented wavelet elements that are allowed to slightly perturb their
locations and orientations to account for the deformations of object shapes.
The model can be easily learned when the objects in the training images
are of
the same pose, and appear at the same location and scale. This is often called
supervised learning. In the situation where the objects may appear
at different unknown locations, orientations and scales in the
training images, we have to incorporate the unknown locations,
orientations and scales as latent variables into the image
generation process, and learn the template by EM-type
algorithms. The E-step imputes the unknown locations, orientations
and scales based on the currently learned template. This step can be
considered self-supervision, which involves using the current
template to recognize the objects in the training images. The M-step
then relearns the template based on the imputed locations,
orientations and scales, and this is essentially the same as
supervised learning. So the EM learning process iterates between
recognition and supervised learning. We illustrate this scheme by
several experiments.
\end{abstract}

%

\end{frontmatter}

\section{Introduction: EM Learning Scheme}

The EM algorithm \cite{DLR} and its variations \cite{Meng} have been
widely used for maximum likelihood estimation of statistical models
when the data are incompletely observed or when there are latent
variables or hidden states. This algorithm is an iterative
computational scheme, where the E-step imputes the missing data or
the latent variables given the currently estimated model, and the
M-step re-estimates the model given the imputed missing data or
latent variables. Besides its simplicity and stability, a key
feature that distinguishes the EM algorithm from other numerical
methods is its interpretability: both the E-step and the M-step
readily admit natural interpretations in a variety of contexts. This
makes the EM algorithm rich, meaningful and inspiring.

In this review article we shall focus on one important context
where the EM algorithm is useful and meaningful, that is, learning
patterns from signals in the settings that are not fully supervised.
In this context, the E-step can be
interpreted as carrying out the
recognition task using the currently learned model of the pattern.
The M-step can be interpreted as relearning the pattern in the
supervised setting, which can often be easily accomplished.

This EM learning scheme has been used in both speech and vision. In
speech recognition, the training of the hidden Markov model
\cite{Rabiner} involves the imputation of the hidden states in the
E-step by the forward and backward algorithms. The M-step computes
the transition and emission frequencies. In computer vision, we want
to learn models for different categories of objects, such as horses,
birds, bikes, etc. The learning is often easy when the objects in
the training images are aligned, in the sense that the objects
appear at the same pose, same location and same scale in the
training images, which are defined on a common image lattice that is
the bounding box of the objects. This is often called supervised
learning. However, it is often the case that the objects appear
at different unknown locations, orientations and scales in the
training images. In such a situation, we have to incorporate the
unknown locations, orientations and scales as latent variables in
the image generation process, and use the EM algorithm to learn the
model for the objects. In the EM learning process, the E-step
imputes the unknown location, orientation and scale of the
object in each training image, based on the currently learned model.
This step uses the current model to recognize the object
in each training image, that is, where it is, at what orientation and
scale. The imputation of the latent variables enables us to align
the training images, so that the objects appear at the same
location, orientation and scale. The M-step then relearns the
model from the aligned images by carrying out supervised learning.
So the EM learning process iterates between recognition and
supervised learning. Recognition is the goal of learning the model,
and it serves as the self-supervision step of the learning process.
The EM algorithm has been used by Fergus, Perona and
Zisserman \cite{Perona} in training the constellation model for
objects.

In this article we shall illustrate EM learning or EM-like learning
by training an active basis model~\cite{Wu,Wu1} that we have recently
developed for deformable templates \cite{Amit,Yuille} of object
shapes. In this model, a template is represented by a linear
composition of a set of localized, elongated and oriented wavelet
elements at selected locations, scales and orientations, and these
wavelet elements are allowed to slightly perturb their locations and
orientations to account for the shape deformations of the objects.
In the supervised setting, the active basis model can be learned by
a shared sketch algorithm, which selects the wavelet elements
sequentially. When a wavelet element is selected, it is shared by
all the training images, in the sense that a perturbed version of
this element seeks to sketch a local edge segment in each training
image. In the situations where learning is not fully supervised, the
learning of the active basis model can be accomplished by the EM-type
algorithms. The E-step recognizes the object in each training image by
matching the image with the currently learned active basis template.
This enables us to align the images. The M-step then relearns the
template from the aligned images by the
shared sketch algorithm.

We would like to point out that the EM algorithm for learning the
active basis model is different than the traditional EM
algorithm, where the model structure is fixed and only the
parameters need to be estimated. In our implementation of the EM
algorithm, the M-step needs to select the wavelet elements in
addition to estimating the parameters associated with the selected
\mbox{elements}. Both the selection of the elements and the estimation of
the parameters are accomplished by maximizing or increasing the
complete-data log-likelihood. So the EM algorithm is used for
estimating both the model structure and the associated parameters.

Readers who wish to learn more about the active basis model are
referred to our recent paper \cite{Wu1}, which is written for the
computer vision community. Compared to that paper, this
review paper is written for the statistical community. In this paper we
introduce the active basis model from an algorithmic perspective,
starting from the familiar problem of variable selection in linear
regression. This paper also provides more details about the EM-type
algorithms than \cite{Wu1}. We wish to convey to the statistical
audience that the problem of vision in general and object recognition
in particular is essentially a statistical problem. We even hope that
this article may attract some statisticians to work on this interesting
but challenging problem.

Section~\ref{sec2} introduces the active basis model for representing
deformable templates, and describes the shared sketch algorithm for
supervised learning. Section~\ref{sec3} presents the EM algorithm for
learning the active basis model in the settings that are not
fully supervised. Section~\ref{sec4} concludes with a discussion.

\section{Active Basis Model: An Algorithmic Tour}\label{sec2}

The active basis model is a natural generalization of the wavelet
regression model. In this section we first explain the\vadjust{\goodbreak}
background and
motivation for the active basis model.
Then we work through a series of
variable selection algorithms for wavelet regression, where the active
basis model emerges naturally.

\subsection{From Wavelet Regression to Active Basis}

\subsubsection{$p>n$ regression and variable section}

Wave\-lets have proven to be immensely useful for signal analysis and
representation \cite{Donohocompression}. Various dictionaries of
wavelets have been designed for different types of signals or
function spaces \cite{CandesDonoho,Simoncellishiftable}. Two key
factors underlying the successes of wavelets are the sparsity of the
representation and the efficiency of the analysis. Specifically, a
signal can typically be represented by a linear superposition of a
small number of wavelet elements selected from an appropriate
dictionary. The selection can be accomplished by efficient
algorithms such as matching pursuit \cite{MallatZhang} and basis
pursuit \cite{Chenbasispursuit}.

%
\begin{figure}

\includegraphics{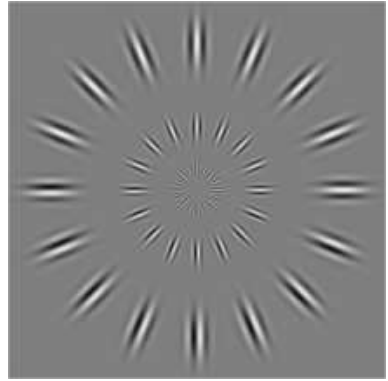}

\caption{A collection of Gabor wavelets at different locations,
orientations and scales. Each Gabor wavelet element is a sine or cosine
wave multiplied by an elongated and oriented Gaussian function. The
wave propagates along the shorter axis of the Gaussian function.}
\label{fig:gf}
\end{figure}

From a linear regression perspective, a signal can be considered a
response variable, and the wavelet elements in the dictionary can be
considered the predictor variables or regressors. The number of
elements in a dictionary can often be much greater than the
dimensionality of the signal, so this is the so-called ``$p > n$''
problem. The selection of the wavelet elements is the variable
selection problem in linear regression. The matching pursuit algorithm
\cite{MallatZhang} is the forward selection method, and the basis
pursuit \cite{Chenbasispursuit} is the lasso method \cite{lasso}.

\subsubsection{Gabor wavelets and simple V1 cells}

Interestingly, wavelet sparse coding also appears to be employed by
the biological visual system for representing natural images. By
assuming the sparsity of the linear representation, Olshausen and
Field \cite{OlshausenField} were able to learn from natural images
a dictionary of localized, elongated, and oriented basis functions that
resemble the Gabor wavelets. Similar wavelets were also
obtained by independent component analysis of natural images
\cite{BellSejnowski}. From a linear regression perspective, Olshausen
and Field essentially asked the following question: Given a sample of
response vectors (i.e., natural images), can we find a dictionary of
predictor vectors or regressors (i.e., basis functions or basis
elements), so that each response vector can be represented as a linear
combination of a small number of regressors selected from the
dictionary? Of course, for different response vectors, different sets
of regressors may be selected from the dictionary.

Figure~\ref{fig:gf} displays a collection
of Gabor wavelet elements at different locations, orientations and
scales. These are sine and cosine waves multiplied by elongated and\vadjust{\goodbreak}
oriented Gaussian functions, where the waves propagate along the
shorter axes of the Gaussian functions. Such Gabor wavelets have been
proposed as mathematical models for the receptive fields of the simple
cells of the primary visual cortex or V1 \cite{Daugman}.

%
\begin{figure*}

\includegraphics{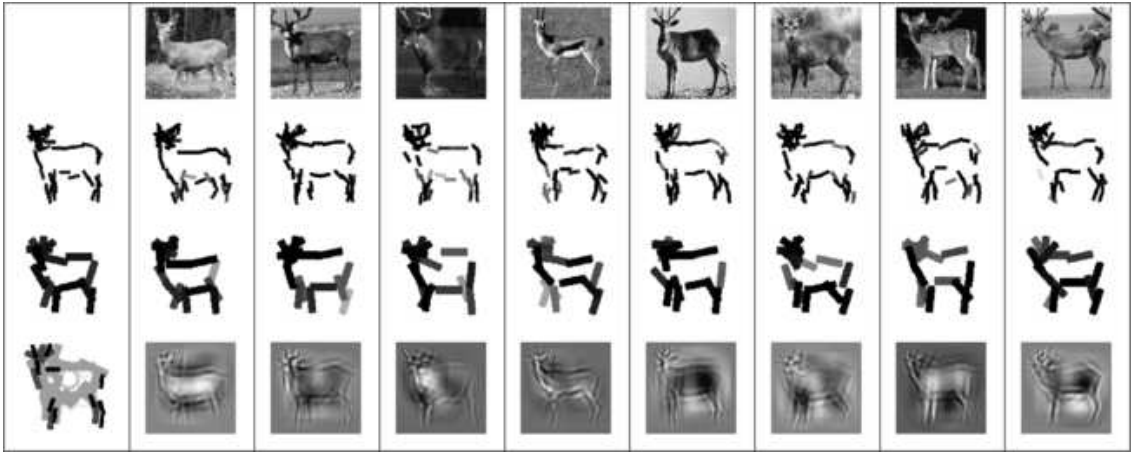}

\caption{Active basis templates. Each Gabor wavelet element is
illustrated by a bar of the same length and at the same location and
orientation as the corresponding element. The first row displays the
training images. The second row displays the templates composed of
50 Gabor wavelet elements at a fixed scale, where the first template is
the common deformable template, and the other templates are deformed
templates for coding the corresponding images. The third row displays
the templates composed of 15 Gabor wavelet elements at a  scale about twice as large as 
those in the second row. In the last row, the template is composed of
wavelet elements at multiple scales, where larger Gabor elements are
illustrated by bars of lighter shades. The rest of the images are
reconstructed by linear superpositions of the wavelet elements of the
deformed templates.}\label{fig:ab}
\end{figure*}

The dictionary of all the Gabor wavelet elements can be very large,
because at each pixel of the image domain, there can be many Gabor
wavelet elements tuned to different scales and orientations. According
to Olshausen and Field \cite{OlshausenField}, the biological visual
system represents a natural image by a linear superposition of a small
number of Gabor wavelet elements selected from such a dictionary.

\subsubsection{From generic classes to specific categories}

Wavelets are designed for generic function classes or learned from
generic ensembles such as natural images, under the generic
principle of sparsity. While such generality offers enormous scope
for the applicability of wavelets, sparsity alone is clearly
inadequate for modeling specific patterns. Recently, we have
developed an active basis model for images of various object classes
\cite{Wu,Wu1}. The model is a natural consequence of seeking a common
wavelet representation simultaneously for multiple training images from
the same object category.

%
\begin{figure*}

\includegraphics{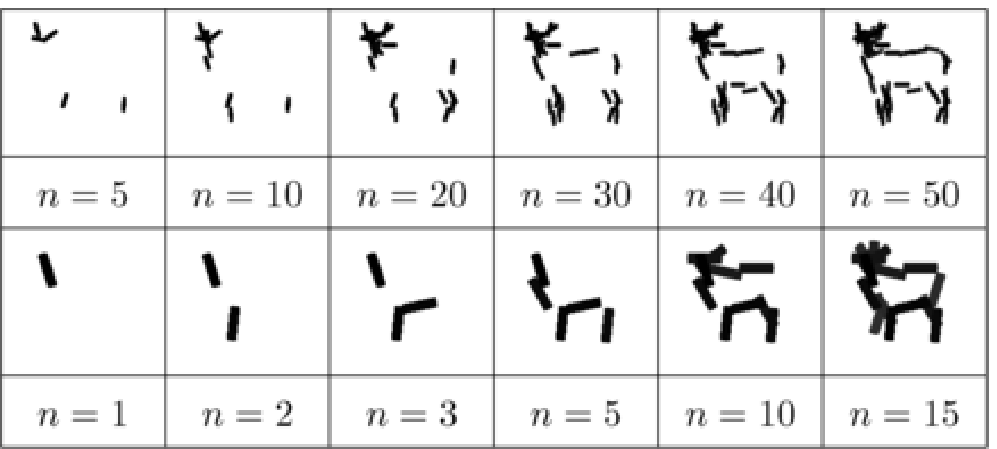}

\caption{Shared sketch process for learning the active basis templates
at two different scales.} \label{fig:ss}
\end{figure*}

The active basis can be learned by the shared sketch algorithm that we
have recently developed \cite{Wu,Wu1}. This algorithm can be
considered a paralleled version of the matching pursuit algorithm \cite{MallatZhang}.
It can also be considered a modification of the
projection pursuit algorithm \cite{Friedman}. The algorithm selects
the wavelet elements sequentially from the dictionary. Each time when
an element is selected, it is shared by all the training images in the
sense that a perturbed version of this element is included in the
linear representation of each image. Figure~\ref{fig:ss}
illustrates the shared sketch process for obtaining the templates
displayed in the second and third rows of Figure~\ref{fig:ab}.

Figure~\ref{fig:ab} illustrates the basic idea. In the first row
there are 8 images of deer. The images are of the same size of 122${}\times{}$120 pixels.
The deer appear at the same location, scale and
pose in these images. For these very similar images,
we want to seek a common wavelet representation, instead of coding
each image individually. Specifically, we want these images to be
represented by
similar sets of wavelet elements, with similar coefficients. We can
achieve this by selecting a common set of wavelet elements, while\vadjust{\goodbreak}
allowing these wavelet elements to locally perturb their locations
and orientations before they are linearly combined to code each
individual image. The perturbations are introduced to account for shape
deformations in the deer. The linear basis formed by such perturbable
wavelet elements is called an active basis.

This is illustrated by the second and third rows of Figure~\ref{fig:ab}.
In each row the first plot displays the common set
of Gabor wavelet elements selected from a dictionary. The dictionary
consists of Gabor wavelets at all the locations and orientations, but
at a fixed scale. Each Gabor wavelet element is symbolically
illustrated by a bar at the same location and orientation and with the
same length as the corresponding Gabor wavelet. So the active basis
formed by the selected Gabor wavelet elements can be interpreted as a
template, as if each element is a stroke for sketching the template.
The templates in the second and third rows are learned using
dictionaries of Gabor wavelets at two different scales, with the scale
of the third row about twice as large as the
scale of the second row. The number of Gabor wavelet elements of the
template in the second row is 50, while the number of elements of
the template in the third row is 15. Currently, we treat this number
as a tuning parameter, although they can be determined in a more
principled way.

Within each of the second and third rows, and for each training
image, we plot the Gabor wavelet elements that are actually used to
represent the corresponding image. These elements are perturbed
versions of the corresponding elements in the first column. So the
templates in the first column are deformable templates, and the
templates in the remaining columns are deformed templates. Thus, the
goal of seeking a common wavelet representation for images from the
same object category leads us to formulate the active basis, which is a
deformable template for the images from the object category.

In the last row of Figure~\ref{fig:ab}, the common template is
learned by selecting from a dictionary that consists of Gabor wavelet
elements at multiple scales instead of a fixed scale. In addition to
Gabor wavelet elements, we also include the center-surround difference
of Gaussian wavelet elements in the dictionary. Such isotropic wavelet
elements are of large scales, and they mainly capture the regional
contrasts in the images. In the template in the last row,
the number of selected wavelet elements is 50.
Larger Gabor
wavelet elements are illustrated by bars of lighter shades. The
difference of Gaussian elements are illustrated by circles. The
remaining images are reconstructed by such multi-scale wavelet
representations, where each image is a linear superposition of the
Gabor and difference of Gaussian wavelet elements of the corresponding
deformed templates.

%
\begin{figure}[b]

\includegraphics{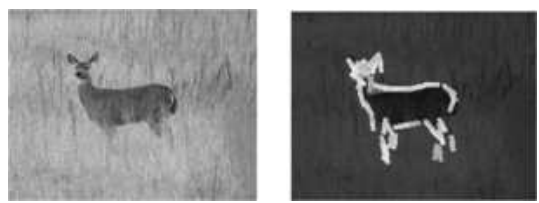}

\caption{Left: Testing image.
Right: Object is detected and sketched
by the deformed template.}\label{fig:find}
\end{figure}
While selecting the wavelet elements of the active basis, we also
estimate the distributions of their coefficients from the training
images. This gives us a statistical model for the images. After
learning this model, we can then use it to recognize the same type of
objects in testing images. See Figure~\ref{fig:find} for an example.
The image on the left is the observed testing image. We scan the
learned template of deer over this image, and at each location, we
match the template to the image by deforming the learned template. The
template matching is scored by the log-likelihood of the statistical
model. We also scan the template over multiple resolutions of the image
to account for the unknown scale of the object in the image. Then we
choose the resolution and location of the image with the maximum
likelihood score, and superpose on the image the deformed template
matched to the image, as shown by the image on the right in Figure~\ref{fig:find}.
This process can be accomplished by a cortex-like
architecture of sum maps and max maps, to be described in Section~\ref{sect:sum-max}.
In machine learning and computer vision
literature, detecting or classifying objects using the learned model is
often called inference. The inference algorithm is often a part of the
learning algorithm. For the active basis model, both learning and
inference can be formulated as maximum likelihood estimation problems.

\subsubsection{Local maximum pooling and complex V1 cells}

Besides wavelet sparse coding theory for V1 simple cells, another inspiration to the active basis
model also comes from
neuroscience. Riesenhuber and Poggio \cite{RiesenhuberPoggio}
observed that the complex cells of the primary visual cortex or V1
appear to perform local maximum pooling of the responses from
simple cells. From the perspective of the active basis model, this
corresponds to estimating the perturbations of the wavelet elements of
the active basis template, so that the template is deformed to match
the observed image. Therefore, if we are to believe Olshausen and
Field's theory on wavelet sparse coding \cite{OlshausenField} and
Riesenhuber and Poggio's theory on local maximum pooling, then the
active basis model seems to be a very natural logical consequence.

In the following subsections we shall describe in detail wavelet sparse
coding, the active basis model, and the learning and inference algorithms.

\subsection{An Overcomplete Dictionary of Gabor Wavelets}

The Gabor wavelets are translated, rotated and dilated versions of
the following function:
\begin{eqnarray*}
G(x_1, x_2) \propto\exp\{ -[(x_1/\sigma_1)^2 + (x_2/\sigma_2)^2]/2\} e^{i {x_1}},
\end{eqnarray*}
which is sine--cosine wave multiplied by a Gaussian function. The
Gaussian function is elongated along the $x_2$-axis, with $\sigma_2 >
\sigma_1$, and the sine--cosine wave propagates along the shorter
$x_1$-axis. We truncate the function to make it locally supported on a
finite rectangular domain, so that it has a well defined length and width.

We then translate, rotate and dilate $G(x_1, x_2)$ to obtain a general
form of the Gabor wavelets:
\begin{eqnarray*}
B_{x_1, x_2, s, \alpha}(x_1', x_2') = G({\tilde{x}_1}/{s},
{\tilde{x}_2}/{s})/s^2,
\end{eqnarray*}
where
\begin{eqnarray*}
\tilde{x_1} &=&(x_1'-x_1)\cos\alpha+(x_2'-x_2)\sin\alpha, \\
\tilde{x_2} &=& -(x_1'-x_1)\sin\alpha+(x_2'-x_2)\cos\alpha.
\end{eqnarray*}
Writing $x = (x_1, x_2)$, each $B_{x, s, \alpha}$ is a localized
function, where $x = (x_1, x_2)$ is the central location, $s$ is the
scale parameter, and $\alpha$ is the orientation. The frequency of the
wave propagation in $B_{x, s, \alpha}$ is $\omega= 1/s$. $B_{x, s,
\alpha}$ = ($B_{x, s, \alpha, 0}$, $B_{x, s, \alpha, 1}$), where
$B_{x, s, \alpha, 0}$ is the even-symmetric Gabor cosine component,
and $B_{x, s, \alpha, 1}$ is the odd-symmetric Gabor sine component.
We always use Gabor wavelets as pairs of cosine and sine components. We
normalize both the Gabor sine and cosine components to have zero mean
and unit $\ell_2$ norm. For each $B_{x, s, \alpha}$, the pair $B_{x,
s, \alpha, 0}$ and $B_{x, s, \alpha, 1}$ are orthogonal to each other.

The dictionary of Gabor wavelets is
\[
\Omega= \{B_{x, s, \alpha}, \forall(x, s, \alpha)\}.
\]
We can discretize the orientation so that $\alpha\in\{o \pi/O,\break o =
0,\ldots, O-1\}$, that is, $O$ equally spaced orientations (the default
value of $O$ is 15 in our experiments). In this article we mostly learn
the active basis template at a fixed scale $s$. The dictionary $\Omega
$ is called ``overcomplete'' because the number of wavelet elements in
$\Omega$ is larger than the number of pixels in the image domain,
since at each pixel, there can be many wavelet elements tuned to
different orientations and scales.

For an image $\mathbf{I}(x)$, with $x \in D$, where $D$ is a set of pixels,
such as a rectangular grid, we can project it onto a Gabor wavelet
$B_{x, s, \alpha, \eta}$, $\eta= 0, 1$. The projection of $\mathbf{I}$ onto
$B_{x, s, \alpha, \eta}$, or the Gabor filter response at $(x, s,
\alpha)$, is
\begin{eqnarray*}
\langle\mathbf{I}, B_{x, s, \alpha, \eta} \rangle= \sum_{x'}
\mathbf{I}(x')
B_{x, s, \alpha, \eta}(x').
\end{eqnarray*}
The summation is over the finite support of $B_{x, s, \alpha, \eta}$.
We write $\langle\mathbf{I}, B_{x, s, \alpha}
\rangle$ = ($\langle\mathbf{I}, B_{x, s, \alpha, 0} \rangle$,
$\langle
\mathbf{I}, B_{x, s, \alpha, 1} \rangle$). The local energy is
\begin{eqnarray*}
|\langle
\mathbf{I}, B_{x, s, \alpha} \rangle|^2 = \langle\mathbf{I}, B_{x, s,
\alpha, 0} \rangle^2+\langle\mathbf{I}, B_{x, s, \alpha, 1}
\rangle^2.
\end{eqnarray*}
$|\langle\mathbf{I}, B_{x, s, \alpha} \rangle|^2$ is the local
spectrum or
the magnitude of the local wave in image $\mathbf{I}$ at $(x, s,
\alpha)$.

Let
\begin{eqnarray*}
\sigma^2_s = \frac{1}{|D|O} \sum_{\alpha} \sum_{x \in D}
|\langle\mathbf{I}, B_{x, s, \alpha} \rangle|^2, \label{eq:normalize}
\end{eqnarray*}
where $|D|$ is the number of pixels in $\mathbf{I}$, and $O$ is the total
number of orientations. For each image $\mathbf{I}$, we normalize it
to $\mathbf{I}
\leftarrow\mathbf{I}/\sigma_s$, so that different images are comparable.

\subsection{Matching Pursuit Algorithm}

For an image $\mathbf{I}(x)$ where $x \in D$, we seek to represent it by
%
\begin{eqnarray}
\mathbf{I}= \sum_{i=1}^{n} c_i B_{x_i, s, \alpha_i} + U, \label{eq:sc}
\end{eqnarray}
where $(B_{x_i, s, \alpha_i}, i = 1,\ldots, n) \subset\Omega$ is a set
of Gabor wavelet elements selected from the dictionary $\Omega$,\break $c_i$
is the coefficient, and $U$ is the unexplained residual image. Recall
that each $B_{x_i, s, \alpha_i}$ is a pair of\break Gabor cosine and sine
components. So $B_{x_i, s, \alpha_i} = (B_{x_i, s, \alpha_i, 0},
B_{x_i, s, \alpha_i, 1})$, $c_{i} = (c_{i, 0}, c_{i, 1})$, and $c_{i}
B_{x_i, s, \alpha_i} = c_{i, 0}B_{x_i, s, \alpha_i, 0} + c_{i, 1}
B_{x_i, s, \alpha_i, 1}$. We fix the scale parameter $s$.

In the representation (\ref{eq:sc}), $n$ is often assumed to be small,
for example, $n = 50$. So the representation (\ref{eq:sc}) is called
sparse representation or sparse coding. This
representation translates a raw intensity image with a huge number
of pixels into a sketch with only a small number of strokes
represented by $\mathbf{B}= (B_{x_i, s, \alpha_i}, i = 1,\ldots, n)$. Because
of the sparsity, $\mathbf{B}$ captures the most visually meaningful elements
in the image. The set of wavelet elements $\mathbf{B}= (B_{x_i, s,
\alpha_i},
i = 1,\ldots, n)$ can be selected from $\Omega$ by the matching pursuit
algorithm \cite{MallatZhang}, which seeks to minimize $\|\mathbf{I}-
\sum
_{i=1}^{n} c_i B_{x_i, s, \alpha_i}\|^2$ by a greedy scheme.
\setcounter{algorithm}{-1}
\begin{algorithm}[(Matching pursuit algorithm)]\label{alg0}
\begin{longlist}[3.]
\item[0.] Initialize $i \leftarrow0$, $U \leftarrow\mathbf{I}$.
\item[1.] Let $i \leftarrow i + 1$. Let $(x_i, \alpha_i) = \arg\max
_{x, \alpha}|\langle U,\break B_{x, s, \alpha}\rangle|^2$.
\item[2.] Let $c_i = \langle U, B_{x_i, s, \alpha_i}\rangle$. Update
$U \leftarrow U - c_i\times\break B_{x_i, s, \alpha_i}$.
\item[3.] Stop if $i = n$, else go back to 1.
\end{longlist}
\end{algorithm}

In the above algorithm, it is possible that a wavelet element is
selected more than once, but this is extremely rare for real images. As
to the choice of $n$ or the stopping criterion, we can stop the
algorithm if $|c_i|$ is below a threshold.

Readers who are familiar with the so-called ``large $p$ and small $n$''
problem in linear regression may have recognized that wavelet sparse
coding is a special case of this problem, where $\mathbf{I}$ is the response
vector, and each $B_{x, s, \alpha} \in\Omega$ is a predictor vector.
The matching pursuit algorithm is actually the forward selection
procedure for variable selection.

The forward selection algorithm in general can be too greedy. But for
image representation, each Gabor wavelet element only explains away a
small part of the image data, and we usually pursue the elements at a
fixed scale, so such a forward selection procedure is not very greedy in this context.

\subsection{Matching Pursuit for Multiple Images}

Let $\{\mathbf{I}_m, m = 1,\ldots, M\}$ be a set of training images
defined on
a common rectangle lattice $D$, and let us suppose that these images
come from the same object category, where the objects appear at the
same pose, location and scale in these images. We can model these
images by a common set of Gabor wavelet elements,
%
\begin{eqnarray}\label{eq:sc1}
\hspace*{15pt}\mathbf{I}_m = \sum_{i=1}^{n} c_{m, i} B_{x_i, s, \alpha_i} + U_m,\quad
 m = 1,\ldots, M.
\end{eqnarray}
$\mathbf{B}= (B_{x_i, s, \alpha_i}, i = 1,\ldots, n)$ can be considered a
common template for these training images. Model (\ref{eq:sc1}) is an
extension of model (\ref{eq:sc}).

We can select these elements by applying the matching pursuit algorithm
on these multiple images simultaneously. The algorithm seeks to
minimize\break $\sum_{m=1}^{M}\|\mathbf{I}_m - \sum_{i=1}^{n} c_{m, i}
B_{x_i, s, \alpha_i}\|^2$ by a greedy scheme.
\begin{algorithm}[(Matching pursuit on multiple images)]\label{alg1}
\begin{longlist}[0.]
\item[0.] Initialize $i \leftarrow0$. For $m = 1,\ldots, M$,
initialize\break
$U_m \leftarrow\mathbf{I}_m$.
\item[1.] $i \leftarrow i + 1$. Select
\[
(x_i, \alpha_i) = \arg\max_{x, \alpha} \sum_{m=1}^{M} |\langle
U_m, B_{x, s, \alpha}\rangle|^2.
\]
\item[2.] For $m = 1,\ldots, M$, let $c_{m, i} = \langle U_m, B_{x_i, s,
\alpha_i}\rangle$, and update $U_m \leftarrow U_{m} - c_{m, i}
B_{x_i, s, \alpha_i}$.
\item[3.] Stop if $i = n$, else go back to 1.
\end{longlist}
\end{algorithm}

Algorithm~\ref{alg1} is similar to Algorithm~\ref{alg0}. The difference is that, in Step
1, $(x_i, \alpha_i)$ is selected by maximizing the sum of the squared
responses.

\subsection{Active Basis and Local Maximum Pooling}

The objects in the training images share similar shapes, but there can
still be considerable variations in their shapes. In order to account
for the shape deformations, we introduce the perturbations to the
common template, and the model becomes
%
\begin{eqnarray}\label{eq:sc2}
\mathbf{I}_m = \sum_{i=1}^{n} c_{m, i} B_{x_i+\Delta x_{m, i}, s,
\alpha_i + \Delta\alpha_{m, i}} + U_m,\nonumber
\\[-8pt]\\[-8pt]
\eqntext{m = 1,\ldots, M.}
\end{eqnarray}
Again, $\mathbf{B}= (B_{x_i, s, \alpha_i}, i = 1,\ldots, n)$ can be considered
a common template for the training images, but this time, this template
is deformable. Specifically, for each image $\mathbf{I}_m$, the wavelet
element $B_{x_i, s, \alpha_i}$ is perturbed to $B_{x_i+\Delta x_{m,
i}, s, \alpha_i + \Delta\alpha_{m, i}}$, where $\Delta x_{m, i}$ is
the perturbation in location, and $\Delta\alpha_{m, i}$ is the
perturbation in orientation. $\mathbf{B}_{m} = (B_{x_i+\Delta x_{m,
i}, s,
\alpha_i + \Delta\alpha_{m, i}}, i = 1,\ldots, n)$ can be considered
the deformed template for coding image $\mathbf{I}_m$. We call the basis
formed by $\mathbf{B}= (B_{x_i, s, \alpha_i}, i = 1,\ldots, n)$ the active
basis, and we call $(\Delta x_{m, i}, \Delta\alpha_{m, i}, i = 1,\ldots, n)$
the activities or perturbations of the basis elements for
image $m$. Model (\ref{eq:sc2}) is an extension of model (\ref{eq:sc1}).

Figure~\ref{fig:ab} illustrates three examples of active basis
templates. In the second and third rows the\break templates in the first
column are $\mathbf{B}= (B_{x_i, s, \alpha_i}, i =\break 1,\ldots, n)$. The scale
parameter $s$ in the second row is smaller than the $s$ in the third
row. For each row, the templates in the remain columns are the deformed
templates $\mathbf{B}_{m} = (B_{x_i+\Delta x_{m, i}, s, \alpha_i
+\Delta\alpha_{m, i}}, i = 1,\ldots, n)$, for $m = 1,\ldots, 8$. The template in
the last row should be more precisely represented by $\mathbf{B}= (B_{x_i,
s_i, \alpha_i}, i = 1,\ldots, n)$, where each element has its own $s_i$
automatically selected together with $(x_i, \alpha_i)$. In this
article we focus on the situation where we fix $s$ (default length of
the wavelet element is 17 pixels).

For the activity or perturbation of a wavelet element $B_{x, s, \alpha}$, we assume that $\Delta x = (d\cos\alpha, d \sin\alpha)$, with
$d \in[-b_1, b_1]$. That is, we allow $B_{x, s, \alpha}$ to shift its
location along its normal direction. We also assume $\Delta\alpha\in
[-b_2, b_2]$. $b_1$ and $b_2$ are the bounds for the allowed
displacements in location and orientation (default values: $b_1 = 6$
pixels, and $b_2 = \pi/15$).
We define
\begin{eqnarray*}
A(\alpha) &=& \bigl\{ \bigl(\Delta x = (d\cos\alpha, d\sin\alpha),
\Delta\alpha\bigr){}\dvtx{}
\\
&&\hspace*{4pt}{}d \in[-b_1, b_1], \Delta\alpha\in[-b_2, b_2]\bigr\}
\end{eqnarray*}
the set of all possible activities for a basis element tuned to
orientation $\alpha$.

We can continue to apply the matching pursuit algorithm to the multiple
training images, the only difference is that we add a local maximum
pooling operation in Steps 1 and 2. The following algorithm is a greedy
procedure to minimize the least squares criterion:
%
\begin{eqnarray}\label{eq:LS}
\sum_{m=1}^{M} \Bigg\|\mathbf{I}_m - \sum_{i=1}^{n} c_{m, i}
B_{x_i+\Delta x_{m,i}, s, \alpha_i + \Delta\alpha_{m, i}}\Bigg\|^2.
\end{eqnarray}

\begin{algorithm}[(Matching pursuit with local maximum pooling)]\label{alg2}
\begin{longlist}[0.]
\item[0.] Initialize $i \leftarrow0$. For $m = 1,\ldots, M$,
initialize\break
$U_m \leftarrow\mathbf{I}_m$.
\item[1.] $i \leftarrow i + 1$. Select
\begin{eqnarray*}
&&\hspace*{-3pt}(x_i, \alpha_i)
\\
&&\hspace*{-3pt}\quad= \arg\max_{x, \alpha} \sum_{m=1}^{M}
\max_{(\Delta x, \Delta\alpha)\in A(\alpha)} |\langle U_m, B_{x+\Delta x, s, \alpha+\Delta\alpha}\rangle|^2.
\end{eqnarray*}
\item[2.] For $m = 1,\ldots, M$, retrieve
\begin{eqnarray*}
&&(\Delta x_{m, i}, \Delta\alpha_{m, i})
\\
&&\quad= \arg\max_{(\Delta x,
\Delta\alpha)\in A(\alpha_i)} |\langle U_m, B_{x_i+\Delta x, s,
\alpha_i+\Delta\alpha}\rangle|^2.
\end{eqnarray*}
Let $c_{m, i} \leftarrow\langle U_m, B_{x_i+\Delta x_{m, i}, s, \alpha
_i+\Delta\alpha_{m, i}}\rangle$, and update $U_m \leftarrow$ $U_m -
c_{m, i} B_{x_i+\Delta x_{m, i}, s, \alpha_i+\Delta\alpha_{m, i}}$.
\item[3.] Stop if $i = n$, else go back to 1.
\end{longlist}
\end{algorithm}

Algorithm~\ref{alg2}\hspace*{-0.01pt} is similar to Algorithm~\ref{alg1}.
The difference is that we add an
extra local maximization operation in Step 1: $\max_{(\Delta x, \Delta
\alpha)\in A(\alpha)} |\langle U_m, B_{x+\Delta x, s, \alpha+\Delta
\alpha}\rangle|^2$. With $(x_i, \alpha_i)$ selected in Step 1, Step
2 retrieves the corresponding maximal $(\Delta x, \Delta\alpha)$ for
each image.
\setcounter{algorithm}{0}
\renewcommand{\thealgorithm}{\arabic{section}.{\arabic{algorithm}}}

We can rewrite Algorithm~\ref{alg2} by defining $R_{m}(x, \alpha) = \langle
U_m, B_{x, s, \alpha}\rangle$. Then instead of updating the residual
image $U_m$ in Step 2, we can update the responses $R_{m}(x, \alpha)$.
\begin{algorithm}[(Matching pursuit with local maximum
pooling)]\label{alg2.1}
\begin{longlist}[2.]
\item[0.] Initialize $i \leftarrow0$. For $m = 1,\ldots, M$,
initalize\break
$R_m(x, \alpha) \leftarrow\langle\mathbf{I}_m, B_{x, s, \alpha
}\rangle$
for all $(x, \alpha)$.
\item[1.] $i \leftarrow i + 1$.
Select
\begin{eqnarray*}
&&(x_i, \alpha_i)
\\
&&\quad= \arg\max_{x, \alpha} \sum_{m=1}^{M}
\max_{(\Delta x, \Delta\alpha)\in A(\alpha)} |R_m(x+\Delta x,
\\
&&\hspace*{156pt}\alpha+\Delta\alpha)|^2.
\end{eqnarray*}
\item[2.] For $m = 1,\ldots, M$, retrieve
\begin{eqnarray*}
&&(\Delta x_{m, i}, \Delta\alpha_{m, i})
\\
&&\quad= \arg\max_{(\Delta x,
\Delta\alpha)\in A(\alpha_i)} |R_m(x_i+\Delta x, \alpha_i+\Delta
\alpha)|^2.
\end{eqnarray*}
Let $c_{m, i} \leftarrow R_m(x_i+\Delta x_{m, i}, \alpha_i+\Delta
\alpha_{m, i})$, and update
\begin{eqnarray*}
&&R_m(x, \alpha)
\\
&&\quad\leftarrow R_m(x, \alpha)- c_{m, i} \langle B_{x, s,
\alpha}, B_{x_i+\Delta x_{m, i}, s, \alpha_i+\Delta\alpha_{m,
i}}\rangle.
\end{eqnarray*}
\item[3.] Stop if $i = n$, else go back to 1.
\end{longlist}
\end{algorithm}

\subsection{Shared Sketch Algorithm}

\setcounter{algorithm}{2}
\renewcommand{\thealgorithm}{\arabic{algorithm}}

Finally, we come to the shared sketch algorithm that we actually used
in the experiments in this paper. The algorithm involves two
modifications to Algorithm~\ref{alg2.1}.\vadjust{\goodbreak}
\begin{algorithm}[(Shared sketch algorithm)]\label{alg3}
\begin{longlist}[0.]
\item[0.] Initialize $i \leftarrow0$. For $m = 1,\ldots, M$,
initialize\break
$R_m(x, \alpha) \leftarrow\langle\mathbf{I}_m, B_{x, s, \alpha
}\rangle$
for all $(x, \alpha)$.
\item[1.] $i \leftarrow i + 1$.
Select
\begin{eqnarray*}
&&(x_i, \alpha_i)
\\
&&\quad= \arg\max_{x, \alpha} \sum_{m=1}^{M}
\max_{(\Delta x, \Delta\alpha)\in A(\alpha)} h\bigl(|R_m(x+\Delta x,
\\
&&\qquad\hspace*{145pt}
\alpha+\Delta\alpha)|^2\bigr).
\end{eqnarray*}
\item[2.] For $m = 1,\ldots, M$, retrieve
\begin{eqnarray*}
&&(\Delta x_{m, i}, \Delta\alpha_{m, i})
\\
&& \quad= \arg\max_{(\Delta x,
\Delta\alpha)\in A(\alpha_i)} |R_m(x_i+\Delta x,
\alpha_i+\Delta \alpha)|^2.
\end{eqnarray*}
Let $c_{m, i} \leftarrow R_m(x_i+\Delta x_{m, i}, \alpha_i+\Delta
\alpha_{m, i})$, and update $R_m(x, \alpha) \leftarrow$ 0 if
\[
\operatorname{corr}(B_{x, s, \alpha}, B_{x_i+\Delta x_{m, i}, s, \alpha
_i+\Delta\alpha_{m, i}})>0.
\]
\item[3.] Stop if $i = n$, else go back to 1.
\end{longlist}
\end{algorithm}

The two modifications are as follows:
\begin{longlist}[(1)]
\item[(1)] In Step 1, we change $|R_m(x+\Delta x, \alpha+\Delta\alpha)|^2$
to $h(|R_m(x+\Delta x, \alpha+\Delta\alpha)|^2)$ where $h(\cdot)$ is a
sigmoid function, which increases from 0 to a saturation level $\xi$
(default: $\xi= 6$),
%
\begin{eqnarray}\label{eq:sigmoid}
h(r) = \xi\biggl[ \frac{2}{1 + e^{-2r/\xi}} - 1 \biggr].
\end{eqnarray}
Intuitively, $\sum_{m=1}^{M} \max_{(\Delta x, \Delta\alpha)\in
A(\alpha)} h(|R_m(x+\Delta x,\break \alpha+\Delta\alpha)|^2)$ can be
considered the sum of the votes from all the images for the location
and orientation $(x, \alpha)$, where each image contributes
$\max_{(\Delta x, \Delta\alpha)\in A(\alpha)}\break h(| R_m(x+\Delta x, \alpha
+\Delta\alpha)|^2)$. The sigmoid transformation prevents a small
number of images from contributing very large values. As a result, the
selection of $(x, \alpha)$ is a more ``democratic'' choice than in
Algorithm~\ref{alg2}, and the selected element seeks to sketch as many edges in
the training images as possible. In the next section we shall formally
justify the use of sigmoid transformation by a statistical model.

\item[(2)] In Step 2, we update $R_m(x, \alpha) \leftarrow 0$ if $B_{x, s,
\alpha}$ is not orthogonal to $B_{x_i+\Delta x_{m, i}, s, \alpha
_i+\Delta\alpha_{m, i}}$. That is,\break we enforce the orthogonality of
the basis
$\mathbf{B}_{m} =\break (B_{x_i+\Delta x_{m, i}, s,
\alpha_i +\Delta\alpha_{m, i}}, i = 1,\ldots, n)$
for each training image $m$. Our
experience with matching pursuit is that it usually selects elements
that have little overlap with each other. So for computational
convenience, we simply enforce that the selected elements are
orthogonal to each other. For two Gabor wavelets $B_1$ and $B_2$, we
define their correlation as $\operatorname{corr}(B_1, B_2) = \sum_{\eta_1 =
0}^{1}\sum_{\eta_2 = 0}^{1} \langle B_{1, \eta_1}, B_{2, \eta
_2}\rangle^2$, that is, the sum of\break squared inner products between the
sine and cosine components of $B_1$ and $B_2$. In practical
implementation, we allow small correlations between selected elements,
that is, we update $R_m(x, \alpha) \leftarrow0$ if $\operatorname{corr}(B_{x,
s, \alpha}, B_{x_i+\Delta x_{m, i}, s, \alpha_i+\Delta\alpha_{m,
i}})>\varepsilon$ (the default value of $\varepsilon=0.1$).
\end{longlist}

\subsection{Statistical Modeling of Images}

In this subsection we develop a statistical model for $\mathbf{I}_m$. A
statistical model is not only important for justifying Algorithm~\ref{alg3} for
learning the active basis template, it also enables us to use the
learned template to recognize the objects in testing images, because we
can use the log-likelihood to score the matching between the learned
template and the image data.

The statistical model is based on the decomposition $\mathbf{I}_m =
\sum
_{i=1}^{m} c_{m, i} B_{x_i+\Delta x_{m, i}, s, \alpha_i+\Delta\alpha
_{m, i}} + U_{m}$, where $\mathbf{B}_{m} = (B_{x_i+\Delta x_{m, i}, s,
\alpha
_i+\Delta\alpha_{m, i}}, i = 1,\ldots, n)$ is orthogonal, and $c_{m, i}
= \langle\mathbf{I}_m, B_{x_i+\Delta x_{m, i}, s, \alpha_i+\Delta
\alpha
_{m, i}}\rangle$, so $U_m$ lives in the subspace that is orthogonal to
$\mathbf{B}_m$. In order to specify a statistical model for $\mathbf{I}_m$ given $\mathbf{B}
_m$, we only need to specify the distribution of $(c_{m, i}, i = 1,\ldots, n)$ and the conditional distribution of $U_m$ given $(c_{m, i}, i
= 1,\ldots, n)$.

The least squares criterion (\ref{eq:LS}) that drives Algorithm~\ref{alg2}
implicitly assumes that $U_{m}$ is white noise, and $c_{m, i}$ follows
a flat prior distribution. These assumptions are wrong. There can be
occasional strong edges in the background, but a white noise $U_m$
cannot account for strong edges. The distribution of $c_{m, i}$ should
be estimated from the training images, instead of being assumed to be a
flat distribution.

In this work we choose to estimate the distribution of $c_{m, i}$ from
the training images by fitting an exponential family model to the
sample $\{c_{m, i}, m = 1,\ldots, M\}$ obtained from the training images,
and we assume that the conditional distribution of $U_m$ given $(c_{m,
i}, i = 1,\ldots, n)$ is the same as the corresponding conditional
distribution in the natural images. Such a conditional distribution can
account for occasional strong edges in the background, and it is the
use of such a conditional distribution of $U_m$ as well as the
exponential family model for $c_{m, i}$ that leads to the sigmoid
transformation in Algorithm~\ref{alg3}. Intuitively, a large response
$|R_m(x+\Delta x, \alpha+\Delta\alpha)|^2$ indicates that there can
be an edge at $(x+\Delta x, \alpha+\Delta\alpha)$. Because an edge
can also be accounted for by the distribution of $U_m$ in the natural
images, a large response should not be taken at its face value for
selecting the basis elements. Instead, it should be discounted by a
transformation such as $h(\cdot)$ in Algorithm~\ref{alg3}.

\subsection{Density Substitution and Projection Pursuit}

More specifically, we adopt the density substitution scheme of
projection pursuit \cite{Friedman} to construct a statistical model.
We start from a reference distribution $q(\mathbf{I})$. In this
article we
assume that $q(\mathbf{I})$ is the distribution of all the natural
images. We
do not need to know $q(\mathbf{I})$ explicitly beyond the marginal
distribution $q(c)$ of $c = \langle\mathbf{I}, B_{x, s, \alpha
}\rangle$
under $q(\mathbf{I})$. Because $q(\mathbf{I})$ is stationary and
isotropic, $q(c)$ is
the same for different $(x, \alpha)$. $q(c)$ is a heavy tailed
distribution because there are edges in the natural images. $q(c)$ can
be estimated from the natural images by pooling a histogram of $\{
\langle\mathbf{I}, B_{x, s, \alpha}\rangle, \forall\mathbf{I},
\forall(x, \alpha
)\}$ where $\{\mathbf{I}\}$ is a sample of the natural images.

Given $\mathbf{B}_m = (B_{x_i+\Delta x_{m, i}, s, \alpha_i+\Delta
\alpha_{m,
i}}, i = 1,\ldots, n)$, we modify the reference distribution $q(\mathbf{I}_m)$
to a new distribution $p(\mathbf{I}_m)$ by changing the distributions of
$c_{m, i}$. Let $p_i(c)$ be the distribution of $c_{m, i}$ pooled from
$\{c_{m, i}, m = 1,\ldots, M\}$, which are obtained from the training
images $\{\mathbf{I}_m, m = 1,\ldots, M\}$. Then we change the
distribution of
$c_{m, i}$ from $q(c)$ to $p_i(c)$, for each $i = 1,\ldots, n$, while
keeping the conditional distribution of $U_{m}$ given $(c_{m, i}, i =
1,\ldots, n)$ unchanged. This leads us to
%
\begin{eqnarray}\label{eq:MODEL}
&&\hspace*{10pt}p\bigl(\mathbf{I}_m | \mathbf{B}_m
= (B_{x_i+\Delta x_{m, i}, s, \alpha_i+\Delta\alpha_{m, i}},i = 1,\ldots, n)\bigr)\nonumber
\\[-8pt]\\[-8pt]
&&\hspace*{10pt}\quad= q(\mathbf{I}_m) \prod_{i=1}^{n} \frac{p_i(c_{m,i})}{q(c_{m, i})},\nonumber
\end{eqnarray}
where we assume that $(c_{m, i}, i = 1,\ldots, n)$ are independent under
both $q(\mathbf{I}_m)$ and $p(\mathbf{I}_m | \mathbf{B}_m)$, for
orthogonal $\mathbf{B}_m$. The
conditional distributions of $U_m$ given $(c_{m, i}, i = 1,\ldots, n)$
under $p(\mathbf{I}_m | \mathbf{B}_m)$ and $q(\mathbf{I}_m)$ are
canceled out in $p(\mathbf{I}_m |
\mathbf{B}_m)/q(\mathbf{I}_m)$ because they are the same. The
Jacobians are also the
same and are canceled out. So $p(\mathbf{I}_m | \mathbf{B}_m)/q(\mathbf{I}_m) =
\prod_{i=1}^{n} {p_i(c_{m, i})}/\break{q(c_{m, i})}$.

The following are three perspectives to view mod\-el~(\ref{eq:MODEL}):
\begin{longlist}[(2)]
\item[(1)] Classification: we may consider $q(\mathbf{I})$ as representing the
negative examples, and $\{\mathbf{I}_m\}$ are positive examples. We
want to
find the basis elements $(B_{x_i, s, \alpha_i}, i = 1,\ldots, n)$ so
that the projections $c_{m, i} = \langle\mathbf{I}_m, B_{x_i+\Delta
x_{m, i},
s, \alpha_i+\Delta\alpha_{m, i}}\rangle$ for $i = 1,\ldots, n$
distinguish the positive examples from the negative examples.

\item[(2)] Hypothesis testing: we may consider $q(\mathbf{I})$ as
representing the
null hypothesis, and the observed histograms of $c_{m, i}, i = 1,\ldots,
n$ are the test statistics that are used to reject the null hypothesis.

\item[(3)] Coding: we choose to code $c_{m, i}$ by $p_i(c)$ instead of $q(c)$,
while continuing to code $U_m$ by the conditional distribution of $U_m$
given $(c_{m, i}, i = 1,\ldots, n)$ under $q(\mathbf{I})$.
\end{longlist}

For all the three perspectives, we need to choose $B_{x_i, s, \alpha
_i}$ so that there is big contrast between $p_i(c)$ and $q(c)$. The
shared sketch process can be considered as sequentially flipping
dimensions of $q(\mathbf{I}_m)$ from $q(c)$ to $p_i(c)$ to fit the observed
images. It is essentially a projection pursuit procedure, with an
additional local maximization step for estimating the activities of the
basis elements.

\subsection{Exponential Tilting and Saturation Transformation}

While $p_i(c)$ can be estimated from $\{c_{m, i}, m = 1,\ldots,\break M\}$ by
pooling a histogram, we choose to parametrize $p_i(c)$ with a single
parameter so that it can be estimated from even a single image.

We assume $p_i(c)$ to be the following exponential family model:
%
\begin{eqnarray}\label{eq:exponential}
p(c; \lambda) = \frac{1}{Z(\lambda)} \exp\{\lambda h(r)\} q(c),
\end{eqnarray}
where $\lambda>0$ is the parameter. For $c = (c_0, c_1)$, $r = |c|^2 =
c_0^2 + c_1^2$,
\begin{eqnarray*}
Z(\lambda) = \int\exp\{\lambda h(r)\}q(c)\,dc = \mathrm{E}_{q}[\exp\{
\lambda
h(r)\}]
\end{eqnarray*}
is the normalizing constant. $h(r)$ is a monotone increasing
function. We assume $p_i(c) = p(c; \lambda_i)$,\break which accounts for
the fact that the squared responses $\{|c_{m, i}|^2 = |\langle\mathbf{I}_m,
B_{x_i+\Delta x_{m, i}, s, \alpha_i+\Delta\alpha_{m, i}}\rangle|^2,
m = 1,\ldots, M\}$ in the positive examples are in general larger than
those in the natural images, because $B_{x_i+\Delta x_{m, i}, s, \alpha
_i+\Delta\alpha_{m, i}}$ tends to sketch a local edge segment in each
$\mathbf{I}_m$. As mentioned before, $q(c)$ is estimated by pooling a
histogram from the natural images.

We argue that $h(r)$ should be a saturation transformation in the sense
that as $r \rightarrow\infty$, $h(r)$ approaches a finite number. The
sigmoid transformation in (\ref{eq:sigmoid}) is such a transformation.
The reason for such a transformation is as follows. Let $q(r)$ be the
distribution of $r = |c|^2 = |\langle\mathbf{I}, B\rangle|^2$ under $q(c)$
where $\mathbf{I}\sim q(\mathbf{I})$. We may implicitly model $q(r)$
as a mixture of
$p_{\rm on}(r)$ and $p_{\rm off}(r)$, where $p_{\rm on}$ is the
distribution of $r$ when $B$ is on an edge in $\mathbf{I}$, and
$p_{\rm off}$
is the distribution of $r$ when $B$ is not on an edge in $\mathbf{I}$.
$p_{\rm
on}(r)$ has a much heavier tail than $p_{\rm off}(r)$. Let $q(r) =
(1-\rho_0)p_{\rm off}(r) + \rho_0 p_{\rm on}(r)$, where $\rho_0$ is
the proportion of edges in the natural images. Similarly, let $p_i(r)$
be the distribution of $r = |c|^2$ under $p_i(c)$. We can model $p_i(r)
= (1-\rho_i)p_{\rm off}(r) + \rho_i p_{\rm on}(r)$, where $\rho_i >
\rho_0$, that is, the proportion of edges sketched by the selected
basis element is higher than the proportion of edges in the natural
images. Then, as $r \rightarrow\infty$, $p_i(r)/q(r) \rightarrow\rho
_i/\rho_0$, which is a constant. Therefore, $h(r)$ should saturate as
$r \rightarrow\infty$.

\subsection{Maximum Likelihood Learning and Pursuit Index}

Now we can justify the shared sketch algorithm as a greedy scheme for
maximizing the log-likelihood. With parametrization (\ref{eq:exponential}) for the statistical model (\ref{eq:MODEL}), the
log-likelihood is
%
\begin{eqnarray}\label{eq:loglik}
\hspace*{15pt}&&\sum_{m=1}^{M} \sum_{i=1}^{n} \log\frac{p_i(c_{m, i})}{q(c_{m, i})}\nonumber
\\
&&\quad= \sum_{i=1}^{n} \Biggl[\lambda_i \sum_{m=1}^{M} h(|\langle\mathbf{I}_m,
B_{x_i+\Delta x_{m, i}, s, \alpha_i+\Delta\alpha_{m, i}}\rangle|^2)
\\
&&\qquad\hspace*{121pt}{}- M \log Z(\lambda_i)\Biggr].\nonumber
\end{eqnarray}
We want to estimate the locations and orientations of the elements of
the active basis, $(x_i, \alpha_i, i = 1,\ldots, n)$, the activities of
these elements, $(\Delta x_{m, i}, \Delta\alpha_{m, i}, i = 1,\ldots,
n)$, and the weights, $(\lambda_i, i = 1,\ldots, n)$, by maximizing the
log-likelihood (\ref{eq:loglik}), subject to the constraints that
$\mathbf{B}_m = (B_{x_i+\Delta x_{m, i}, s, \alpha_i+\Delta\alpha_{m, i}}$,
$i= 1,\ldots, n)$ is orthogonal for each $m$.

First, we consider the problem of estimating the weight $\lambda_i$
given $\mathbf{B}_m$. To maximize the log-likelihood (\ref{eq:loglik}) over
$\lambda_i$, we only need to maximize
\begin{eqnarray*}
l_i(\lambda_i)
&=& \lambda_i\sum_{m=1}^{M} h(|\langle\mathbf{I}_m,
B_{x_i+\Delta x_{m, i}, s, \alpha_i+\Delta\alpha_{m, i}}\rangle|^2)
\\
&&{}- M \log Z(\lambda_i).
\end{eqnarray*}
By setting $l_i'(\lambda_i) = 0$, we get the well-known form of the
estimating equation for the exponential family model,
%
\begin{eqnarray}\label{eq:MLe}
&&\mu(\lambda_i)\nonumber
\\[-8pt]\\[-8pt]
&&\quad= \frac{1}{M}\sum_{m=1}^{M} h(|\langle\mathbf{I}_m,
B_{x_i+\Delta x_{m, i}, s, \alpha_i+\Delta\alpha_{m, i}}\rangle|^2),\nonumber
\end{eqnarray}
where the mean parameter $\mu(\lambda)$ of the exponential family
model is
%
\begin{eqnarray}\label{eq:ee}
\mu(\lambda) &=& \mathrm{E}_{\lambda}[h(r)]\nonumber
\\[-8pt]\\[-8pt]
&=& \frac{1}{Z(\lambda)}\int h(r) \exp\{\lambda h(r)\} q(r)\,dr.\nonumber
\end{eqnarray}
The estimating equation (\ref{eq:MLe}) can be solved easily because
$\mu(\lambda)$ is a one-dimensional function. We can simply store
this monotone function over a one-dimen\-sional grid. Then we solve this
equation by looking up the stored values, with the help of nearest
neighbor linear interpolation for the values between the grid points.
For each grid point of $\lambda$, $\mu(\lambda)$ can be computed by
one-dimensional integration as in (\ref{eq:ee}). Thanks to the
independence assumption, we only need to deal with such one-dimensional
functions, which relieves us from time consuming MCMC computations.

Next let us consider the problem of selecting $(x_i, \alpha_i)$, and
estimating the activity $(\Delta x_{m, i}, \Delta\alpha_{m, i})$ for
each image $\mathbf{I}_m$. Let $\hat{\lambda}_i$ be the solution to the
estimat\-ing equation (\ref{eq:MLe}). $l_i(\hat{\lambda}_i)$ is
monotone in $\sum_{m=1}^{M} h(|\langle\mathbf{I}_m,\break B_{x_i+\Delta
x_{m, i},
s, \alpha_i+\Delta\alpha_{m, i}}\rangle|^2)$. Therefore, we need to
find $(x_i, \alpha_i)$, and $(\Delta x_{m, i}, \Delta\alpha_{m,
i})$, by maximizing\break $\sum_{m=1}^{M} h(|\langle\mathbf{I}_m,
B_{x_i+\Delta
x_{m, i}, s, \alpha_i+\Delta\alpha_{m, i}}\rangle|^2)$. This
justifies Step 1 of Algorithm~\ref{alg3}, where $\sum_{m=1}^{M} h(|R_m(x+\Delta
x, \break \alpha+\Delta\alpha)|^2)$ serves as the pursuit index.

\subsection{SUM-MAX Maps for Template Matching}\label{sect:sum-max}

After learning the active basis model, in particular, the basis
elements $\mathbf{B}= (B_{x_i, s, \alpha_i}, i = 1,\ldots, n)$ and the weights
$(\lambda_i, i = 1,\ldots, n)$, we can use the learned model to find the
object in a testing image $\mathbf{I}$, as illustrated by Figure~\ref{fig:find}. The testing image may not be defined on the same lattice
as the training images. For example, the testing image may be larger
than the training images. We assume that there is one object in the
testing image, but we do not know the location of the object in the
testing image. In order to detect the object, we scan the template over
the testing image, and at each location $x$, we can deform the template
and match it to the image patch around $x$. This gives us a
log-likelihood score at each location $x$. Then we can find the maximum
likelihood location $\hat{x}$ that achieves the maximum of the
log-likelihood score among all the $x$. After computing $\hat{x}$, we
can then retrieve the activities of the elements of the active basis
template centered at~$\hat{x}$.
\begin{algorithm}[(Object detection by template\break matching)]\label{alg4}
\begin{longlist}[1.]
\item[1.] For every $x$, compute
\begin{eqnarray*}
&&l(x)
\\
&&\quad= \sum_{i=1}^{n} \Bigl[ \lambda_i \max_{(\Delta x, \Delta
\alpha) \in A(\alpha_i)} h(|\langle\mathbf{I}, B_{x+x_i+\Delta x,
s, \alpha_i + \Delta\alpha}\rangle|^2)
\\
&&\hspace*{177pt}{}- \log Z(\lambda_i) \Bigr].
\end{eqnarray*}

\item[2.] Select $\hat{x} = \arg\max_{x} l(x)$. For $i = 1,\ldots, n$, retrieve
\begin{eqnarray*}
&&(\Delta x_i, \Delta\alpha_i)
\\
&&\quad= \arg\max_{(\Delta x, \Delta\alpha)
\in A(\alpha_i)} |\langle\mathbf{I}, B_{\hat{x}+x_i+\Delta x, s,
\alpha_i+ \Delta\alpha}\rangle|^2.
\end{eqnarray*}
\item[3.] Return the location $\hat{x}$, and the deformed template
$(B_{\hat{x}+x_i+\Delta x_i, s, \alpha_i + \Delta\alpha_i}, i = 1,\ldots, n)$.
\end{longlist}
\end{algorithm}

Figure~\ref{fig:find} displays the deformed template\break $(B_{\hat
{x}+x_i+\Delta x_i, s, \alpha_i + \Delta\alpha_i}, i = 1,\ldots, n)$,
which is superpo\-sed on the image on the right.

Step 1 of the above algorithm can be realized by a computational
architecture called sum-max maps.
\setcounter{algorithm}{0}
\renewcommand{\thealgorithm}{4.{\arabic{algorithm}}}
\begin{algorithm}[(sum-max maps)]\label{alg4.1}
\begin{longlist}[1.]
\item[1.] For all $(x, \alpha)$, compute $\mathrm{SUM}1(x, \alpha) =
h(|\langle\mathbf{I},\break B_{x, s, \alpha}\rangle|^2)$.
\item[2.] For all $(x, \alpha)$, compute
\[
\mathrm{MAX}1(x, \alpha) =
\max_{(\Delta x, \Delta\alpha) \in A(\alpha)}\mathrm{SUM}1(x+\Delta x,
\alpha+\Delta\alpha).
\]
\item[3.] For all $x$, compute $\mathrm{SUM}2(x) = \sum_{i=1}^{n}
[\lambda_i\times\break \mathrm{MAX}1(x+x_i, \alpha_i) - \log Z(\lambda_i)]$.
\end{longlist}
$\mathrm{SUM}2(x)$ is $l(x)$ in Algorithm~\ref{alg4}.
\end{algorithm}

The local maximization operation in Step 2 of Algorithm~\ref{alg4.1} has been
hypothesized as the function of the complex cells of the primary visual
cortex \cite{RiesenhuberPoggio}. In the context of the active basis
model, this operation can be justified as the maximum likelihood
estimation of the activities. The shared sketch learning algorithm can
also be written in terms of sum-max maps.

%
\begin{figure*}[b]
\vspace*{5pt}

\includegraphics{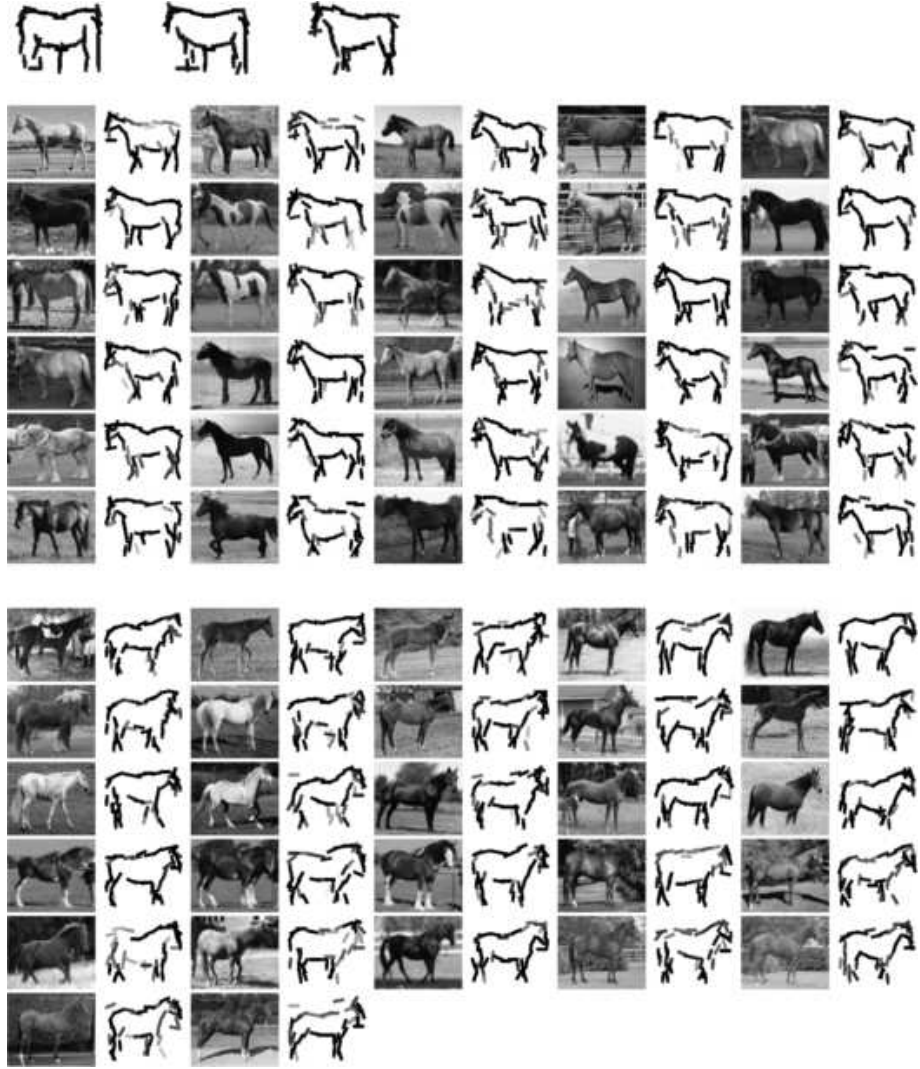}

\caption{Template learned from images of horses facing two different
directions. The first row displays the templates learned in the
first 3 iterations of the EM algorithm. For each training image,
the deformed template is plotted to the right of it. The number of
training images is 57. The image size is $120\times100$ (width${}\times{}$height). The
number of elements is 40. The number of EM iterations is 3. }
\label{fig:horse}
\end{figure*}

\setcounter{algorithm}{4}
\renewcommand{\thealgorithm}{\arabic{algorithm}}

The activities $(\Delta x_{m, i}, \Delta\alpha_{m, i}, i = 1,\ldots,
n)$ should be treated as latent variables in the active basis model.
However, in both learning and inference algorithms, we treat them as
unknown parameters, and we maximize over them instead of integrating
them out. According to Little and Rubin \cite{Little}, maximizing the
complete-data likelihood over the latent variables may not lead to
valid inference in general. However, in natural images, there is little
noise, and the uncertainty in the activities is often very small. So
maximizing over the latent variables can be considered a good
approximation to integrating out the latent variables.

\section{Learning Active Basis Templates by EM-Type Algorithms}\label{sec3}

The shared sketch algorithm in the previous section requires that
the objects in the training images $\{\mathbf{I}_m\}$ are of the same pose,
at the same location and scale, and the lattice of $\mathbf{I}_m$ is the
bounding box of the object in\vadjust{\goodbreak}
$\mathbf{I}_m$. It is often the case
that the
objects may appear at different unknown locations, orientations and
scales in $\{\mathbf{I}_m\}$. The unknown locations, orientations and scales
can be incorporated into the image generation process as hidden
variables. The template can still be learned by the maximum likelihood method.

\subsection{Learning with Unknown Orientations}

We start from a simple example of learning a horse template at the
side view, where the horses can face either to the left or to the
right. Figure~\ref{fig:horse} displays the results of EM learning.
The three templates in the first row are the learned templates in
the first three iterations of the EM algorithm.
The rest of the figure displays the
training images, and for each training image, a deformed template is
displayed to the right of it. The EM algorithm correctly estimates
the direction for each horse, as can be seen by how the algorithm
flips the template to sketch each training image.

Let $\mathbf{B}= (B_{i} = B_{x_i, s, \alpha_i}, i = 1,\ldots, n)$ be the\break
deformable template of the horse, and $\mathbf{B}_m =\break(B_{m, i} =
B_{x_i+\Delta x_{m, i}, s, \alpha_i + \Delta\alpha_{m, i}}$,
$i = 1,\ldots, n)$ be the deformed template for $\mathbf{I}_m$. Then $\mathbf{I}_m$
can either be generated by $\mathbf{B}_m$ or the mirror reflection of $\mathbf{B}_m$,
that is, $(B_{R(x_i+\Delta x_{m, i}), s, -(\alpha_i + \Delta\alpha_{m, i})},
i = 1,\ldots, n)$, where for\break $x = (x_1, x_2)$, $R(x) = (-x_1, x_2)$ (we
assume that the template is centered at origin). We can introduce a
hidden variable $z_m$ to account for this uncertainty, so that $z_m =
1$ if $\mathbf{I}_m$ is generated by $\mathbf{B}_m$, and $z_m = 0$ if
$\mathbf{I}_m$ is
generated by the mirror reflection of $\mathbf{B}_m$. More formally,
we can
define $\mathbf{B}_{m}(z_m)$, so that\break $\mathbf{B}_m(1) = \mathbf{B}_m$, and $\mathbf{B}_m(0)$ is
the mirror reflection of $\mathbf{B}_m$. Then we can assume the following
mixture model: $z_m \sim\operatorname{Bernoulli}(\rho)$, where
$\rho$ is the prior probability that $z_m = 1$, and $[\mathbf{I}_m | z_m]
\sim p(\mathbf{I}_m | \mathbf{B}_m(z_m), \Lambda)$, where $\Lambda=
(\lambda_i, i
= 1,\ldots, n)$. We need to learn $\mathbf{B}$, and estimate $\Lambda$ and
$\rho$.

A simple observation is that $p(\mathbf{I}_m | \mathbf{B}_{m}(z_m)) =\break
p(\mathbf{I}_m(z_m)
| \mathbf{B}_m)$, where $\mathbf{I}_m(1) = \mathbf{I}_m$ and $\mathbf{I}_m(0)$ is the mirror
reflection of $\mathbf{I}_m$. In other words, in the case of $z_m =
1$, we
do not need to make any change to $\mathbf{I}_m$ or $\mathbf{B}_m$.
In the case of
$z_m = 0$, we can either flip the template or flip the image, and
these two alternatives will produce the same value for the likehood
function.

In the EM algorithm, the E-step imputes $z_m$ for $m = 1,\ldots, M$
using the current template $\mathbf{B}$. This means recognizing the
orientation of the object in $\mathbf{I}_m$. Given $z_m$, we can
change $\mathbf{I}
_m$ to $\mathbf{I}_m(z_m)$, so that $\{\mathbf{I}_m(z_m)\}$ become
aligned with each
other, if
$z_m$ are imputed correctly. Then in the M-step, we can learn the template from the aligned
images $\{\mathbf{I}_m(z_m)\}$ by the shared sketch algorithm.

The complete data log-likelihood for the $m$th observation is
\begin{eqnarray*}
&&\log p(\mathbf{I}_m, z_m | \mathbf{B}_m)
\\
&&\quad=z_m \log p(\mathbf{I}_m | \mathbf{B}_m, \Lambda)
\\
&&\qquad{}+ (1-z_m) \log p\bigl(\mathbf{I}_m (0) |\mathbf{B}_m, \Lambda\bigr) \nonumber\\
&&\qquad{}+ z_{m} \log\rho+ (1-z_m) \log(1-\rho),
\end{eqnarray*}
which is linear in $z_m$. So in the E-step, we only need to compute
the predictive expectation of $z_m$,
\begin{eqnarray*}
\hat{z}_m &=& \operatorname{Pr}(z_m = 1 | \mathbf{B}_m, \Lambda, \rho) \nonumber
\\
&=& \frac{\rho p(\mathbf{I}_m | \mathbf{B}_m, \Lambda)}
{\rho p(\mathbf{I}_m | \mathbf{B}_m, \Lambda) + (1-\rho) p(\mathbf{I}_m(0) | \mathbf{B}_m,
\Lambda)}.
\end{eqnarray*}
Both $\log p(\mathbf{I}_m| \mathbf{B}_m, \Lambda)$ and $\log
p(\mathbf{I}_m(0) | \mathbf{B}_m,
\Lambda)$ are\break readily available in the M-step.

%
\begin{figure*}

\includegraphics{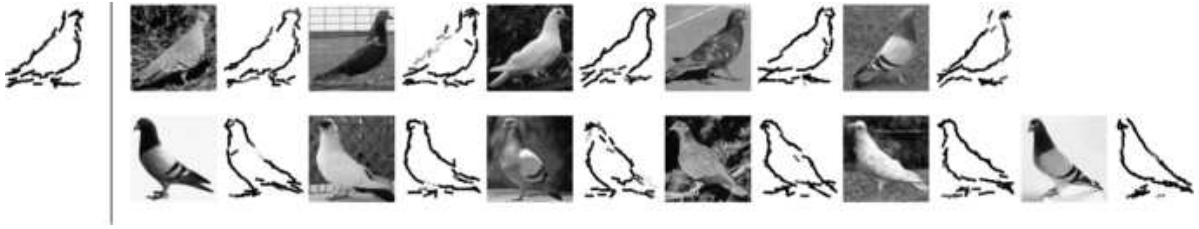}

\caption{Template learned from 11 images of pigeons facing different
directions. The image size is $150\times150$. The number of
elements is 50. The number of iterations is 3.}\label{fig:pigeon}
\end{figure*}

%
\begin{figure*}[b]

\includegraphics{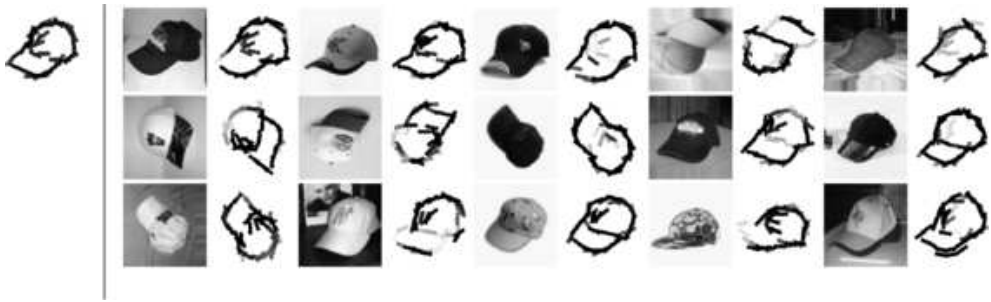}

\caption{Template learned from 15 images of baseball caps facing
different orientations. The image size is $100\times100$. The
number of elements is 40. The number of iterations is 5.}\label{fig:cap}
\end{figure*}

The M-step seeks to maximize the expectation of the complete-data
log-likelihood,
%
\begin{eqnarray}\label{eq:L1}
\hspace*{23pt}&& \sum_{i=1}^{n} \Biggl[ {\lambda}_i \sum_{m=1}^{M}
\bigl(\hat{z}_{m} h(|\langle\mathbf{I}_m, B_{m, i}
\rangle|^2)\nonumber
\\
&&\quad\hspace*{40pt}{}+(1-\hat{z}_{m})h(|\langle\mathbf{I}_m(0), B_{m, i}
\rangle|^2)\bigr)
\\
&&\quad\hspace*{113pt}{}- M \log Z({\lambda}_i)\Biggr]\nonumber
\\\label{eq:L2}
&&\quad{}+ \Biggl[\log\rho\sum_{m=1}^{M} \hat{z}_m\nonumber
\\[-8pt]\\[-8pt]
&&\hspace*{75pt}{}+ \log(1-\rho) \Biggl(M - \sum_{m=1}^{M} \hat{z}_m\Biggr)\Biggr].\nonumber
\end{eqnarray}
The maximization of (\ref{eq:L2}) leads to $\hat{\rho} =
\sum_{m=1}^{M} \hat{z}_m/M$. The maximization of (\ref{eq:L1}) can
be accomplished by the shared sketch algorithm, that is, Algorithm~\ref{alg3},
with the following minor modifications:
\begin{longlist}[(1)]
\item[(1)] The training images become $\{\mathbf{I}_{m}, \mathbf{I}_{m}(0),
m = 1,\ldots,
M\}$, that is, there are $2M$ training images instead of $M$ images.
Each $\mathbf{I}_m$ contributes two copies, the original copy $\mathbf{I}_m$ or
$\mathbf{I}_m(1)$, and the mirror reflection $\mathbf{I}_m(0)$. This
reflects the
uncertainty in $z_m$. For each image $\mathbf{I}_m$, we attach a weight
$\hat{z}_m$ to $\mathbf{I}_m$, and a weight $1-\hat{z}_m$ to
$\mathbf{I}_m(0)$.
Intuitively, a fraction of the horse in $\mathbf{I}_m$ is at the same
orientation as the current template, and a fraction of it is at the
opposite orientation---a ``Schrodinger horse'' so to speak. We use
$(\mathbf{J}_{k}, w_k, k = 1,\ldots, 2M)$ to represent these $2M$ images and
their weights.
\item[(2)] In Step 1 of the shared sketch algorithm, we select $(x_i, \alpha
_i)$ by
\begin{eqnarray*}
&&(x_i, \alpha_i)
\\
&&\quad= \arg\max_{x, \alpha} \sum_{k=1}^{2M} w_{k}
\max_{(\Delta x, \Delta\alpha)\in A(\alpha)}
h\bigl(|R_k(x+\Delta x,
\\
&&\hspace*{175pt}\alpha+\Delta\alpha)|^2\bigr).
\end{eqnarray*}
\item[(3)] The maximum likelihood estimating equation for
$\lambda_i$ is
\begin{eqnarray*}
&&\mu(\lambda_i)
\\
&&\quad= \frac{1}{M} \sum_{k=1}^{2M} w_k
\max_{(\Delta x, \Delta\alpha)\in A(\alpha_i)} h\bigl(|R_k(x_i+\Delta x,
\\
&&\hspace*{157pt}\alpha_i+\Delta\alpha)|^2\bigr),
\end{eqnarray*}
where the right-hand side is the weighted average obtained from the
$2M$ training
images.
\item[(4)] Along with the selection of $B_{x_i, s, \alpha_i}$ and the
estimation of $\lambda_i$, we should calculate the
template matching scores
\begin{eqnarray*}
&&\hspace*{-5pt}\log p(\mathbf{J}_k | \mathbf{B}_k, \Lambda)
\\
&&\hspace*{-5pt}\quad=
\sum_{i=1}^{n} \Bigl[\hat{\lambda}_i \max_{(\Delta x, \Delta
\alpha)\in A(\alpha_i)}h\bigl(|R_k(x_i+\Delta x,
\\
&&\qquad\hspace*{115pt}{}
\alpha_i+\Delta\alpha)|^2\bigr)
\\
&&\quad\hspace*{125pt}{}- \log Z(\hat{\lambda}_i)\Bigr],
\end{eqnarray*}
for $k = 1,\ldots, 2M$. This gives us $\log p(\mathbf{I}_m| \mathbf{B}_m, \Lambda)$
and $\log p(\mathbf{I}_m(0) | \mathbf{B}_m, \Lambda)$, which can
then be used in the E-step.
\end{longlist}

We initialize the algorithm by randomly generating $\hat{z}_{m} \sim
\operatorname{Unif}[0, 1]$, and then iterate between the M-step and the
E-step. We stop the algorithm after a few iterations. Then we
estimate $z_m = 1$ if $\hat{z}_{m}>1/2$ and $z_m = 0$ otherwise.

In Figure~\ref{fig:horse} the results are obtained after 3
iterations of the EM algorithm. Initially, the learned template is
quite symmetric, reflecting the confusion of the algorithm regarding
the directions of the horses. Then the EM algorithm begins a process
of ``symmetry breaking'' or ``polarization.'' The slight asymmetry
in the initial template will push the algorithm toward favoring for
each image the direction that is consistent with the majority
direction. This process quickly leads to all the images aligned to one
common direction.

Figure~\ref{fig:pigeon} shows another example where a template of
a pigeon is learned from examples with mixed directions.

We can also learn a common template when the objects are at
more than two different orientations in the training images. The
algorithm is essentially the same as described above. Figure~\ref{fig:cap} displays the learning of the
template of a baseball cap from examples where the caps turn to
different orientations. The E-step involves rotating the images by
matching to the current template, and the M-step learns the template
from the rotated images.

\subsection{Learning From Nonaligned Images}

\begin{figure*}

\includegraphics{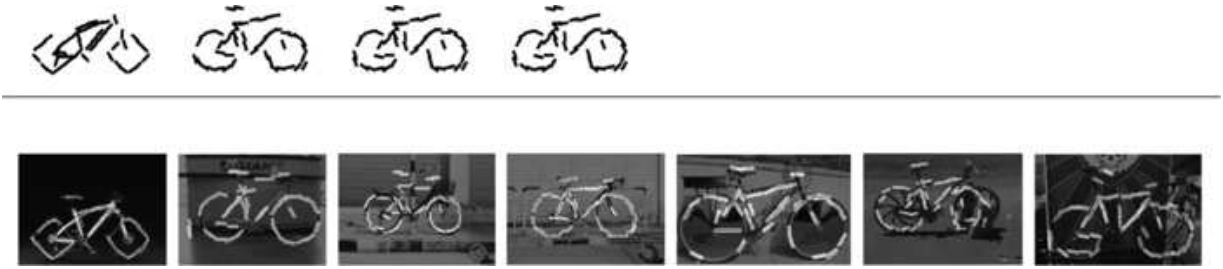}

\caption{The first row shows the sequence of
templates learned in the first 3 iterations. The first one is the
starting template, which is learned from the first training image.
The second row shows the  bikes detected by the learned template, where a
deformed template is superposed on each training image. The size of the
template is $225 \times169$. The number of elements is 50. The
number of iterations is 3.}\label{fig:bike}
\end{figure*}

When the objects appear at different locations in the training
images, we need to infer the unknown locations while
learning the template. Figure~\ref{fig:bike} displays the
template of a bike learned from the 7 training images where the
objects appear at different locations and are not aligned. It also
displays the deformed templates superposed on the objects in the
training images.

%

In order to incorporate the unknown locations into the image
generation process, let us assume that both the learned
template $\mathbf{B}= (B_{x_i, s, \alpha_i}, i = 1,\ldots, n)$ and the training
images $\{\mathbf{I}_m\}$ are centered at
origin. Then let us assume that the location of the object\break in
image $\mathbf{I}_m$ is $x^{(m)}$, which is assumed to be uniformly
distributed within the image lattice of $\mathbf{I}_m$. Let us define
$\mathbf{B}_m(x^{(m)}) = (B_{x^{(m)} + x_i+\Delta x_{m, i}, s, \alpha
_i +\Delta\alpha_{m, i}},\break i = 1,\ldots, n)$ to be the deformed template
obtained by translating the template $\mathbf{B}$ from the origin to $x^{(m)}$
and then deforming it. Then the generative model for $\mathbf{I}_m$ is
$p(\mathbf{I}
_m | \mathbf{B}_m(x^{(m)}), \Lambda)$.

Just like the example of learning the horse template, we can
transfer the transformation of the template to the transformation of
the image data, and the latter transformation leads to the alignment
of the images. Let us define $\mathbf{I}_m(x^{(m)})$ to be the image obtained
by translating the image $\mathbf{I}_m$ so that the center of $\mathbf{I}_m(x^{(m)})$
is $-x^{(m)}$. Then  $p(\mathbf{I}_m | \mathbf{B}_m(x^{(m)}),\break
\Lambda) = p(\mathbf{I}_m(x^{(m)})| \mathbf{B}_m, \Lambda)$.
If we
know $x^{(m)}$ for $m = 1,\ldots, M$, then the images $\{\mathbf{I}_m(x^{(m)})\}
$ are all aligned, so that we can learn a template from these aligned
images by the shared sketch algorithm. On the other hand, if we know
the template, we can use the template to recognize and locate the
object in each $\mathbf{I}_m$ by the inference algorithm, that is, Algorithm~\ref{alg4},
using the sum-max maps, and identify $x^{(m)}$. Such considerations
naturally lead to the iterative EM-type scheme.

The complete-data log-likelihood is
%
\begin{eqnarray}\label{eq:La}
&&\sum_{i=1}^{n} \Biggl[ {\lambda}_i \sum_{m=1}^{M}
h\bigl(\big|\big\langle\mathbf{I}_m(x^{(m)}),\nonumber
\\
&&\hspace*{66pt}B_{x_i+\Delta x_{m, i}, s, \alpha_i
+ \Delta\alpha_{m, i}}\big\rangle\big|^2\bigr)
\\
&&\hspace*{101pt}{}- M \log Z({\lambda}_i)\Biggr].\nonumber
\end{eqnarray}

In the E-step we perform the recognition task by calculating
\begin{eqnarray*}
p_m(x) = \operatorname{Pr}\bigl(x^{(m)} = x | \mathbf{B}, \Lambda\bigr) \propto p\bigl(\mathbf{I}_m(x) | \mathbf{B}_m,
\Lambda\bigr),\quad  \forall x.
\end{eqnarray*}
That is, we scan the template over the whole image $\mathbf{I}_m$, and at
each location $x$, we evaluate the template matching between
the image $\mathbf{I}_m$ and the translated and deformed template
$\mathbf{B}_m(x)$.
$\log p(\mathbf{I}_m(x) | \mathbf{B}_m, \Lambda)$ is the $\mathrm{SUM}2(x)$ output by the
sum-max maps in Algorithm~\ref{alg4.1}. This gives us $p_m(x)$, which is the
posterior or
predictive distribution of the unknown location $x^{(m)}$ within
the image lattice of $\mathbf{I}_m$. We can then use $p_m(x)$ to compute
the expectation of the complete-data log-likelihood (\ref{eq:La}) in
the E-step.

Our experience suggests that $p_m(x)$ is always high\-ly peaked at
a particular position. So instead of computing the average of
(\ref{eq:La}), we simply impute $x^{(m)} = \arg\max_{x} p_m(x)$.

Then in the M-step, we maximize the complete data log-likelihood
(\ref{eq:La}) by the shared sketch\break algorithm, that is, we learn the
template $\mathbf{B}$ from $\{\mathbf{I}_m(x^{(m)})\}$. This step performs
supervised learning from the aligned images.

In our current experiment we initialize the algorithm by learning
$(\mathbf{B}
, \Lambda)$ from the first image. In learning from this single
image, we set $b_1 = b_2 = 0$, that is, we do not allow the
elements $(B_{x_i, s, \alpha_i}, i = 1,\ldots, n)$ to perturb. After
that, we reset $b_1$ and $b_2$ to
their default values, and iterate the recognition step and the
supervised learning step.

In addition to the unknown locations, we also allow the uncertainty
in scales. In the recognition step, for each $\mathbf{I}_m$, we search
over a
number of different resolutions of $\mathbf{I}_m$. We take $\mathbf{I}_m(x^{(m)})$ to
be the optimal resolution that contains the maximum template matching
score across all the resolutions.

\begin{figure*}

\includegraphics{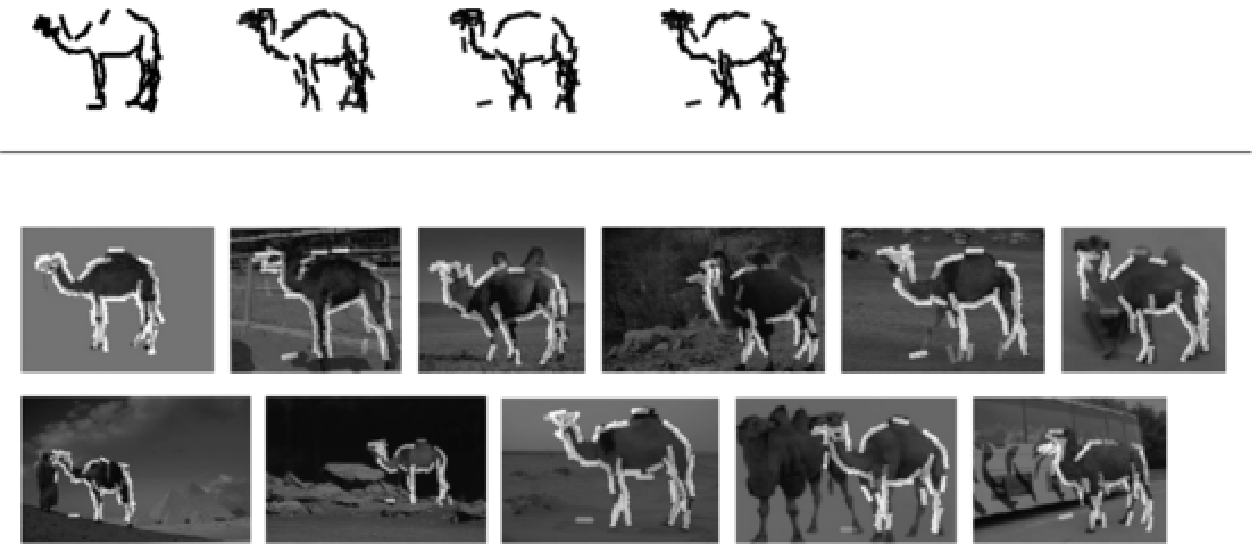}

\caption{The first row shows the sequence of
templates learned in iterations 0, 1, 3, 5. The second and third
rows show
the camel images with superposed deformed templates. The size of the
template is $192\times145$. The number of elements is 60. The number
of iterations is 5.} \label{fig:camel}
\end{figure*}

\begin{figure*}

\includegraphics{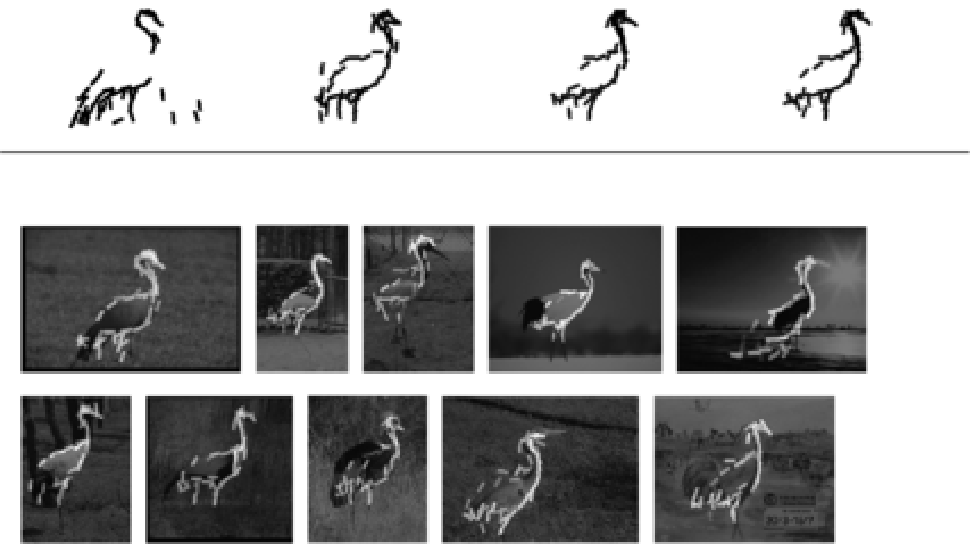}

\caption{The first row shows the sequence of templates learned in
iterations 0, 1, 3, 5. The other rows show
the crane images with superposed deformed templates. The
size of the template is $285\times190$. The number of elements is
50. The number of iterations is 5.} \label{fig:crane}
\end{figure*}

\begin{figure*}

\includegraphics{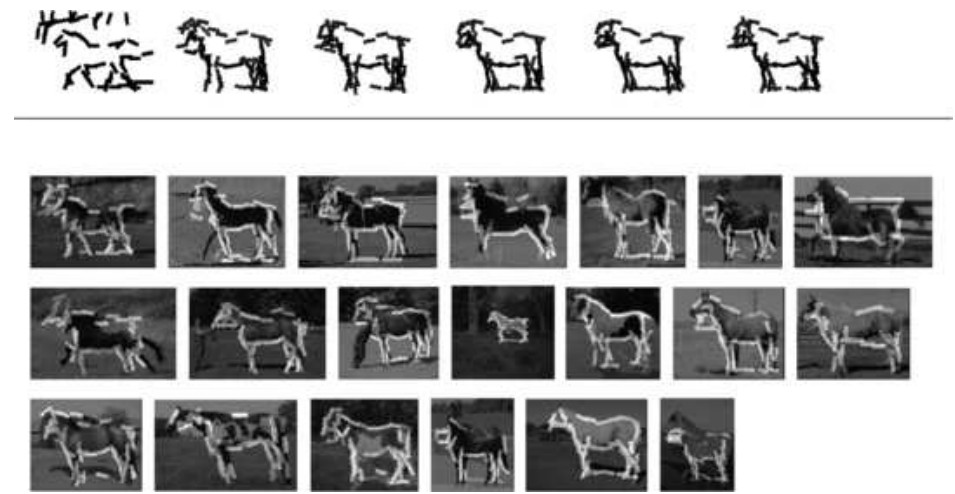}

\caption{The first row shows the sequence of
templates learned in  iterations 0, 2, 4, 6, 8, 10.
The other rows show
the horse images with superposed deformed templates. We use the
first 20 images of the Weizmann horse data set \protect\cite{Borenstein},
which are resized to half the original sizes.
The size of the template is 158${}\times{}$116.
The number of elements is 60.
The number of iterations is 10. The detection results on the
rest of the images in this data set can be found in the reproducibility
page. } \label{fig:Whorse}
\end{figure*}

\begin{figure*}

\includegraphics{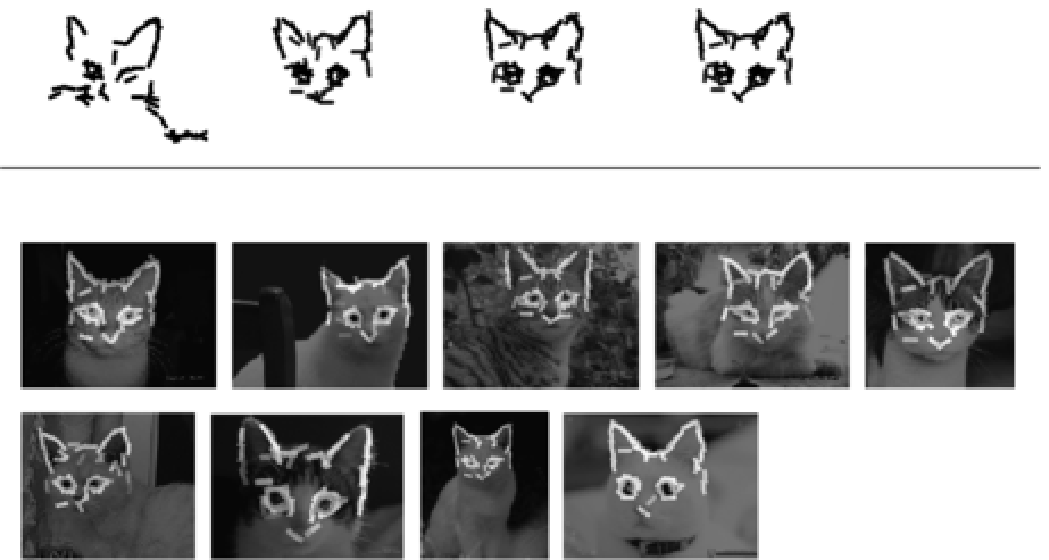}

\caption{The size of the template is $240\times180$. The number of
elements is 60. The number of iterations is 3.} \label{fig:cat}
\end{figure*}

In Figure~\ref{fig:bike} the first row displays the templates
learned over the EM iterations. The first template is learned from the
first training image. Figures \ref{fig:camel}--\ref{fig:cat}
display more examples.

%

%

%

%

\section{Discussion}\label{sec4}

This paper experiments with EM-type algorithms for learning active
basis models from training images where the objects may appear at
unknown locations, orientations and scales. For more details on
implementing the shared sketch algorithm, the reader is referred to
\cite{Wu1} and the source code posted on the reproducibility page.

We would like to emphasize two aspects of the algorithms that are
different from the usual EM algorithm. The first aspect is that the
M-step involves the selection of the basis elements, in addition to the
estimation of the associated parameters. The second aspect is that the
performance of the algorithms can rely heavily on the
initializations. In learning from nonaligned images, the
algorithm is initialized by training the active basis model on a
single image. Because of the simplicity of the model, it is
possible to learn the model from a single image. In addition, the
learning algorithm seems to converge within a~few iterations.

\subsection{Limitations}

The active basis model is a simple extension of the wavelet
representation. It is still very limited in the following aspects.
The model cannot account for large deformations, articulate
shapes, big changes in poses and view points, and occlusions. The
current form of the model does not describe textures and lighting
variations either. The current version of the learning algorithm
only deal with situations where there is one object in each
image. Also, we have tuned two parameters in our implementation.
One is the image resize factor that we apply to the training
images before the model is learned. Of course, for each experiment, a
single resize factor is applied to all the training images. The other
parameter is the number of elements in the active basis.

\subsection{Possible Extensions}

It is possible to extend the active basis model to address some of
the above limitations. We shall discuss two directions of
extensions. One is to use active basis models as parts of the
objects. The other is to train active basis models by local
learning.

\textit{Active basis models as part-templates:} The active basis model
is a composition of a number of Gabor wavelet elements. We can
further compose multiple active basis models to represent more
articulate shapes or to account for large deformations by allowing
these active basis models to change their overall locations, scales
and orientations within limited ranges. These active basis models
serve as part-templates of the whole composite template. This is
essentially a hierarchical recursive compositional structure
\cite{Gemancomposition,Zhugrammar}. The inference or template
matching can be based on a recursive structure of sum-max maps.
Learning such a structure should be possible by extending the
learning \mbox{algorithms} studied in this article. See \cite{Wu1} for
preliminary results. See also \cite{Ross,Sudderth} for recent work on
part-based models.

\textit{Local learning of multiple prototype templates:} In each
experiment we assume that all the training images share a common
template. In reality, the training images may contain different
types of objects, or different poses of the same type of objects. It
is therefore necessary to learn multiple prototype templates. It is
possible to do so by modifying the current learning algorithm. After
initializing the algorithm by single image training, in the M-step,
we can relearn the template only from the $K$ images with the
highest template matching scores, that is, we relearn the template
from the $K$ nearest neighbors of the current template. Such a
scheme is consistent with the EM-clustering algorithm for fitting
mixture models. We can start the algorithm from every training
image, so that we learn a local prototype template around each
training image. Then we can trim and merge these prototypes. See \cite{Wu1} for preliminary results.

\section*{Reproducibility}

All the experimental results reported in this paper can be reproduced by the code
that we have~posted at
\url{http://www.stat.ucla.edu/\textasciitilde ywu/ActiveBasis.html}.

\section*{Acknowledgments}
We are grateful to the editor of the special
issue and the two reviewers for their valuable comments that have
helped us improve the presentation of the paper. The work reported in
this paper has been supported by NSF Grants DMS-07-07055, DMS-10-07889, IIS-07-13652,
ONR N00014-05-01-0543, Air Force Grant FA 9550-08-1-0489 and the Keck foundation.

\end{document}